\documentclass[sigconf, authorversion]{acmart}

\AtBeginDocument{%
  \providecommand\BibTeX{{%
    \normalfont B\kern-0.5em{\scshape i\kern-0.25em b}\kern-0.8em\TeX}}}

\setcopyright{acmcopyright}
\copyrightyear{2021}
\acmYear{2021}
\acmDOI{10.1145/xxxxxxx.xxxxxxx}

\acmConference[WiSec '21]{WiSec '21: 14th ACM Conference on Security and Privacy in Wireless and Mobile Networks}{June 28--July 1, 2021}{Abu Dhabi, United Arab Emirates}
\acmBooktitle{WiSec '21: 14th ACM Conference on Security and Privacy in Wireless and Mobile Networks, June 28--July 1, 2021, Abu Dhabi, United Arab Emirates}
\acmPrice{15.00}
\acmISBN{978-1-4503-XXXX-X/21/06}

\usepackage{enumitem}

\begin{document}

\title{LNGate: Powering IoT with Next Generation Lightning Micro-payments using Threshold Cryptography}

\author{Ahmet Kurt}
\affiliation{%
  \institution{Dept. of Elec. and Comp. Engineering\\ Florida International University}
  \streetaddress{10555 W Flagler St.}
  \city{Miami, FL}
  \country{USA}}
\email{akurt005@fiu.edu }

\author{Suat Mercan}
\affiliation{%
  \institution{Dept. of Elec. and Comp. Engineering\\ Florida International University}
  \city{Miami, FL}
  \country{USA}}
\email{smercan@fiu.edu }

\author{Omer Shlomovits}
\affiliation{%
  \institution{ZenGo X}
  \city{Tel Aviv}
  \country{Israel}}
\email{omer@zengo.com}

\author{Enes Erdin}
\affiliation{%
  \institution{Dept. of Computer Science\\ University of Central Arkansas}
  \city{Conway, AR}
  \country{USA}}
\email{eerdin@uca.edu}

\author{Kemal Akkaya}
\affiliation{%
  \institution{Dept. of Elec. and Comp. Engineering\\ Florida International University}
  \streetaddress{10555 W Flagler St.}
  \city{Miami, FL}
  \country{USA}}
\email{kakkaya@fiu.edu }

\renewcommand{\shortauthors}{Kurt et al.}

\begin{abstract}
Bitcoin has emerged as a revolutionary payment system with its decentralized ledger concept however it has significant problems such as high transaction fees and long confirmation times. Lightning Network (LN), which was introduced much later, solves most of these problems with an innovative concept called off-chain payments. With this advancement, Bitcoin has become an attractive venue to perform micro-payments which can also be adopted in many IoT applications (e.g. toll payments). Nevertheless, it is not feasible to host LN and Bitcoin on IoT devices due to the storage, memory, and processing requirements. Therefore, in this paper, we propose an efficient and secure protocol that enables an IoT device to use LN through an untrusted gateway node. The gateway hosts LN and Bitcoin nodes and can open \& close LN channels, send LN payments on behalf of the IoT device. This delegation approach is powered by a (2,2)-threshold scheme that requires the IoT device and the LN gateway to jointly perform all LN operations which in turn secures both parties' funds. Specifically, we propose to thresholdize LN's Bitcoin public and private keys as well as its commitment points. With these and several other protocol level changes, IoT device is protected against revoked state broadcast, collusion, and ransom attacks. We implemented the proposed protocol by changing LN's source code and thoroughly evaluated its performance using a Raspberry Pi. Our evaluation results show that computational and communication delays associated with the protocol are negligible. To the best of our knowledge, this is the first work that implemented threshold cryptography in LN.
\end{abstract}





\acmYear{2021}\copyrightyear{2021}
\setcopyright{acmcopyright}
\acmConference[WiSec '21]{Conference on Security and Privacy in Wireless and Mobile Networks}{June 28--July 2, 2021}{Abu Dhabi, United Arab Emirates}
\acmBooktitle{Conference on Security and Privacy in Wireless and Mobile Networks (WiSec '21), June 28--July 2, 2021, Abu Dhabi, United Arab Emirates}
\acmPrice{15.00}
\acmDOI{10.1145/3448300.3467833}
\acmISBN{978-1-4503-8349-3/21/06}

\maketitle

\section{Introduction}

The last decade has witnessed a fast adoption of IoT in numerous domains as it provides great opportunities and convenience \cite{gubbi2013internet}. These devices have typically been utilized for continuous data collection as they are equipped with various sensors. The improvement in their computational capabilities has enabled applications where an IoT device may pay or get paid for provided/received services. For instance, a vehicle passing through a toll gate can make a payment from its on-board unit (OBU) which acts as an IoT device to communicate with toll infrastructure \cite{pavsalic2016vehicle}. Similarly; electric vehicle charging, parking, sensor data selling, etc. are some other potential applications in this specific area \cite{mercan2021cryptocurrency, kurt2020lnbot}.

Automated payments in device-to-device (D2D) communication without any human intervention is a desired feature. Although it might be possible to link these devices to traditional payment systems such as credit cards, it requires third-party involvement which might bring management overhead as well as privacy concerns. In this aspect, cryptocurrencies play a crucial role to create a more convenient payment system. Thus, combining IoT and cryptocurrency systems such as Bitcoin \cite{nakamoto2019bitcoin} and Ethereum \cite{wood2014ethereum} can help address these challenges.

However; long confirmation times, low throughput, and high transaction fees have been the main concern in mainstream cryptocurrencies which limits their scalability \cite{chauhan2018blockchain} as well as inhibits their adoption for micro-payment scenarios. Thus, payment channel network (PCN) idea was proposed which addresses these problems by utilizing off-chain transactions as a second layer solution \cite{decker2015fast}. As an example, Lightning Network (LN) \cite{poon2016bitcoin} has been introduced as the PCN solution for Bitcoin. LN currently exceeds 20,000 nodes since its emergence in three years. Although LN addresses the problems mentioned above, it still cannot be accommodated on most of the IoT devices because of the computation, communication, and storage requirements \cite{lniotproblem}. Specifically, using LN requires running an LN node along with a full Bitcoin node which in total, occupies more than 340 GB of storage. Robust Internet connection and relatively high computation power are also necessary for Bitcoin block verification.

Therefore, a lightweight solution is needed to enable resource-constrained IoT devices to use LN for micro-payments. To this end, in this paper, we propose a threshold cryptography-based protocol where an IoT device can perform LN operations through an \textit{untrusted LN gateway} that hosts the full LN and Bitcoin nodes. With this integration, the IoT device can 1) open LN channels, 2) send LN payments, and 3) close LN channels. The LN gateway is incentivized to provide this payment service by the service fees it charges for sending IoT device's payments.

In our proposed protocol, we utilize (2,2)-threshold cryptography to enable mutual control of LN channels by the IoT device and the LN gateway. The IoT device and the LN gateway cooperatively authorize channel operations by incorporating their threshold secret shares into signing and key generation processes. This enables the IoT device to get involved in LN operations without the risk of losing its funds. More specifically, the LN gateway cannot spend IoT device's funds in the channel without involving the IoT device in (2,2)-threshold operations. While we are proposing changes to LN, they only need to be applied to LN gateway's LN node which makes the protocol compatible with the rest of the network. This is the beauty of using threshold cryptography. Since LN's original protocol is modified, this necessitates re-visiting the known \textit{revoked state broadcast} issue of LN. We propose solutions to handle revoked state broadcasts by making small modifications to the LN gateway's commitment transaction. These modifications also totally disincentives the LN gateway from colluding with other LN nodes in the network as well as from ransoming \cite{oz2021survey} the IoT device.

To assess the effectiveness and overhead of the proposed protocol, we implemented it by changing the source code of one of the main LN implementations. In our test setup, we used a Raspberry Pi as an IoT device and a server PC running on the cloud as an LN gateway. We demonstrated that the proposed protocol 1) enables realization of timely payments; 2) can be run on networks with low bandwidth (data rate); and 3) associated costs of using the protocol for IoT devices are negligible. In addition to those, we considered a real-life case where a vehicle at a certain speed makes toll payments through a wireless connection. We demonstrated that the proposed protocol enables realization of timely toll payments in the example real-life scenario. We separately provide a security analysis of the proposed protocol.

\vspace{2mm}
\textbf{Contributions}: Our contributions in this work are as follows:
\vspace{-2mm}
\begin{itemize}[leftmargin=*]
    \item We propose LNGate, a secure and lightweight protocol that enables resource-constrained IoT devices to make instant, cheap micro-payments using Bitcoin's Lightning Network. We utilize (2,2)-threshold cryptography so that IoT does not have to get involved in Bitcoin or LN operations, but only in transaction signing and key generation.
    \item We propose to thresholdize LN's Bitcoin public/private keys and commitment points for a secure 2-party threshold LN node which is resistant against revoked state broadcast, collusion, and ransom attacks.
    \item We implemented LNGate by changing LN's source code. LN's Bitcoin public and private keys were thresholdized. Our evaluation results show that LNGate enables fast IoT micro-payments with minimal overhead.
\end{itemize}

The rest of the paper is structured as follows. Section \ref{sec:background} describes the LN, its components, and specifications along with some preliminaries. System and threat model are given in Section \ref{sec:systemmodel}. The proposed protocol is explained in detail in Section \ref{sec:protocol}. Section \ref{sec:securityanalysis} explains how our protocol mitigates the threats mentioned in Section \ref{sec:systemmodel}. Detailed performance evaluation of the proposed protocol is given in Section \ref{sec:evaluation}. In Section \ref{sec:RelatedWork}, we discuss the relevant work. Finally, we conclude the paper in Section \ref{sec:conclusion}.

\vspace{-2mm}
\section{Background}
\label{sec:background}

This section provides background on LN, its underlying mechanisms and threshold cryptography as a preliminary to our proposed approach. 

\vspace{-2mm}
\subsection{Lightning Network}
LN was introduced in 2015 in a draft technical whitepaper \cite{poon2016bitcoin} and later was implemented and deployed onto Bitcoin Mainnet by Lightning Labs \cite{lnlaunch}. It runs on top of the Bitcoin blockchain as a \textit{second-layer} peer-to-peer distributed PCN and aims to address the scalability problem of Bitcoin. It enables opening secure payment channels among users to perform instant and cheap Bitcoin transfers through multi-hop routes within the network by utilizing Bitcoin's smart contract capability \cite{contract}. The number of users using LN has grown significantly since its creation. At the time of writing this paper, LN incorporates 20,085 nodes and 44,088 channels which hold 1275 BTC in total (worth around 75 Million USD) \cite{1ml}.

\begin{figure*}[htb]
    \centering
    \includegraphics[width=0.9\linewidth]{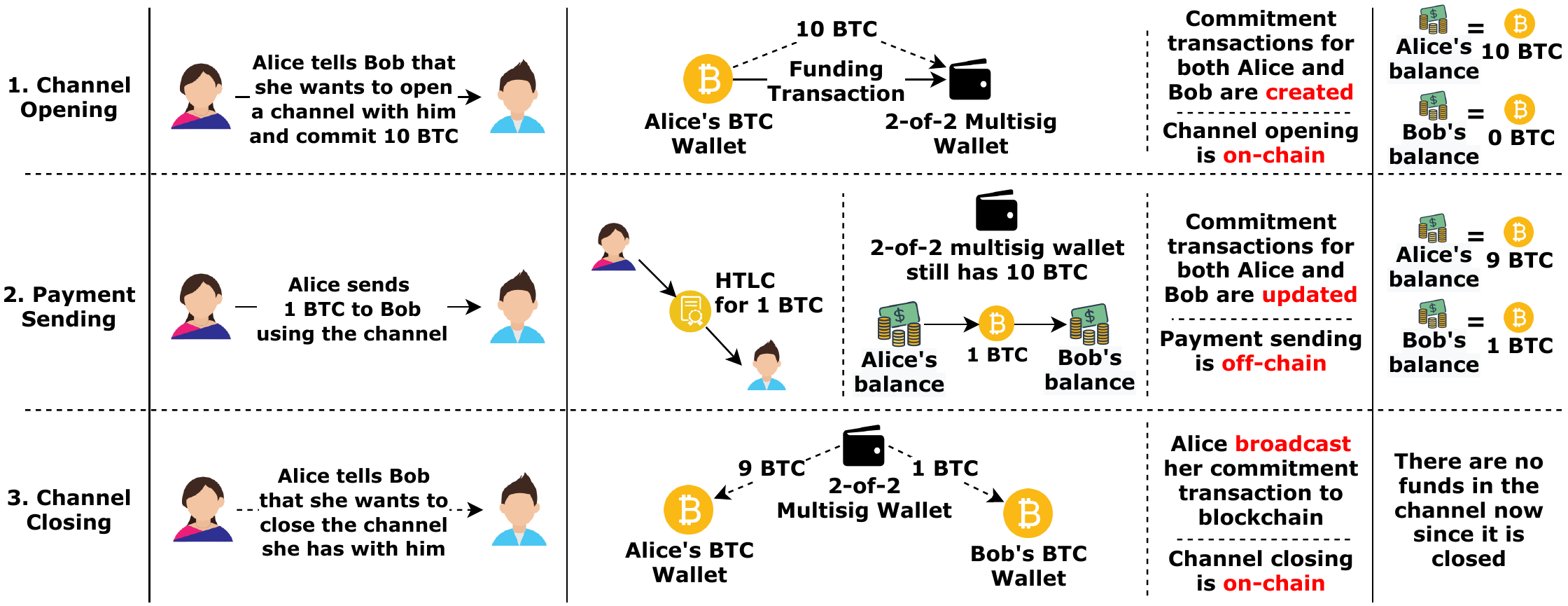}
    \vspace{-4mm}
    \caption{\small Life cycle of an LN channel.}
    \label{fig:background}
    \vspace{-3mm}
\end{figure*}

\vspace{-2mm}
\subsection{Lightning Network Preliminaries}
\label{sec:preliminaries}

This section explains technical concepts such as \textit{funding transaction}, \textit{commitment transaction}, \textit{Hash Time Locked Contract (HTLC)}, and \textit{revoked state} which are essential to understand our protocol. Throughout the explanations, we use the case that Alice wants to send payments to Bob and opens an LN payment channel to him. In a payment channel, the payments can flow in both directions between Alice and Bob to exchange funds without committing the transactions to the Bitcoin blockchain. This means that these transactions take place off-chain.

\vspace{1mm}

\noindent \textit{Funding transaction}: Channels are established by an on-chain Bitcoin transaction called the \textit{funding transaction}. This transaction determines the \textit{capacity} of the channel. Channels are funded by the party opening the channel. For example, if Alice funds the channel with 10 Bitcoins (BTC) as shown in Fig. \ref{fig:background} (1. Channel Opening), she can send payments to Bob until her 10 BTC are exhausted. 

\vspace{1mm}

\noindent \textit{Commitment Transactions}: To keep track of the channel balances of the parties, \textit{commitment transactions} are created mutually. When a new payment is sent, the state of the channel is updated to reflect the new balances of the parties. The balances in each state are held by the commitment transactions. A commitment transaction is essentially a Bitcoin transaction therefore it can be programmed to create complex conditions that parties have to meet to spend the outputs. This is possible with Bitcoin's \textit{scripting system} \cite{script} which we utilize in our protocol.

When Alice creates the funding transaction, she also creates her own and Bob's version of the commitment transactions. She sends her signature for Bob's version of commitment transaction to Bob as well as the funding transaction's \textit{outpoint}\footnote{Outpoint is different from output and a good explanation can be found at \cite{outpoint}.}, which is the combination of the transaction ID and output index number. This funding transaction outpoint is used as an input to the commitment transaction. When Bob receives the outpoint, he generates the signature for Alice's version of commitment transaction and sends it to her. This way, both Alice and Bob have cross-signed commitment transactions that they can use to close the channel if they want. Fig. \ref{fig:background} illustrates these processes.

A typical commitment transaction have three outputs. For Alice's commitment transaction; the first output shows Alice's current balance in the channel which is time-locked, while the second output is for Bob's balance in the channel which is immediately spendable by him. The third output is related to payments as will be explained consequently. 

\vspace{1mm}

\noindent \textit{Revoked State Broadcast}: The first output in a commitment transaction is conditional such that, if Alice closes the channel by broadcasting her commitment transaction, she can redeem her funds from this output only after waiting \textit{k} number of blocks. This is a security mechanism in LN to prevent cheating since the commitment transaction Alice broadcast might be a \textit{revoked} commitment transaction (i.e., an older state). If this is the case, Bob can claim this output using the \textit{revocation private key} and spend it immediately while Alice is waiting, consequently punishing Alice.

\vspace{1mm}

\noindent \textit{HTLC}: When Alice sends 1 BTC payment to Bob as depicted in Fig. \ref{fig:background} (2. Payment Sending), she adds a third output to the commitment transaction which is called the \textit{HTLC output}. In this scheme, Alice asks Bob to generate a secret called \textit{preimage}. Bob hashes the preimage and sends the hash to Alice. Alice includes the hash in the HTLC and sends the HTLC to Bob. Upon receiving Alice's HTLC, Bob reveals the preimage to Alice to receive the 1 BTC payment. This acts as a proof that Bob is the intended recipient. If for some reason, Bob cannot reveal the preimage on time to claim the HTLC, Alice gets the 1 BTC back after some block height \textit{w}.

\vspace{-2mm}
\subsection{Basis of Lightning Technology (BOLT)}
\label{sec:bolts}



Basis of Lightning Technology (BOLT) specifications describe LN's layer-2 protocol for secure off-chain Bitcoin payments. In order to implement our proposed protocol, we made modifications on BOLT \#2 which is LN's peer protocol for channel management. BOLT \#2 has three phases which are \textit{channel establishment}, \textit{normal operation} of the channel, and \textit{channel closing}. Using this protocol, LN nodes talk to each other for channel related operations. More information regarding BOLT \#2 is provided at Appendix \ref{sec:appendixbolt2}.

\vspace{-2mm}
\subsection{Threshold Cryptography}
\label{sec:thresholdbackground}

\textit{Threshold cryptography} \cite{desmedt1994threshold} is a subset of Secure Multiparty Computation (SMPC) and deals with cryptographic operations where more than one party is involved. In a cryptosystem, secrets also known as private keys need to be protected from stealing. This is especially important for cryptocurrencies, as the compromisation of private keys will result in funds being stolen. Therefore, the idea of sharing a private key among several trusted parties was suggested \cite{shamir1979share}. In a threshold scheme, a secret is shared among $n$ parties and a threshold $t$ is defined such that, no group of $t-1$ can learn anything about the secret. Such setup is defined as $(t, n)$-threshold scheme. We utilize (2,2)-threshold cryptography in our proposed protocol for cooperative signing and key generation. Bitcoin uses Elliptic Curve Digital Signature Algorithm (ECDSA) for generating its public and private keys as well as for signing. Since LN is working on top of Bitcoin, it also uses ECDSA. In the next section, we briefly explain the ECDSA signature scheme. The details of the 2-party ECDSA threshold key generation and signing that we employ in our protocol are given at Appendix \ref{appendix:keygeneration} and \ref{appendix:signing}. They are adapted from \cite{gothamcitywhitepaper} and we use parts of Lindell's work \cite{lindell2017fast}.

\vspace{-2mm}
\subsection{ECDSA Signature Scheme}
\label{sec:ECDSA}

Having a private key $x$, and a corresponding public key $x \cdot G$; Alice wants to sign a message $m$. She can compute an ECDSA signature on $m$ by following the steps below:

\begin{enumerate}[leftmargin=*]
    \item An ephemeral key $k$ satisfying $k \in [1,q]$ is chosen.
    \item $R=k \cdot G$ is computed.
    \item An $r$ is set such that it is the $x$ coordinate of the curve point $R$.
    \item Signature $s$ is computed which equals to $s = k^{-1} \cdot (H(m) + r \cdot x)$. $H$ is a hash function (i.e. SHA256).
    \item $(r,s)$ is the ECDSA signature.
\end{enumerate}

\vspace{-2mm}
\section{System \& Threat Model}
\label{sec:systemmodel}

\subsection{System Model}

\begin{figure}[h]
    \centering
    \vspace{-2mm}
    \includegraphics[width=\linewidth]{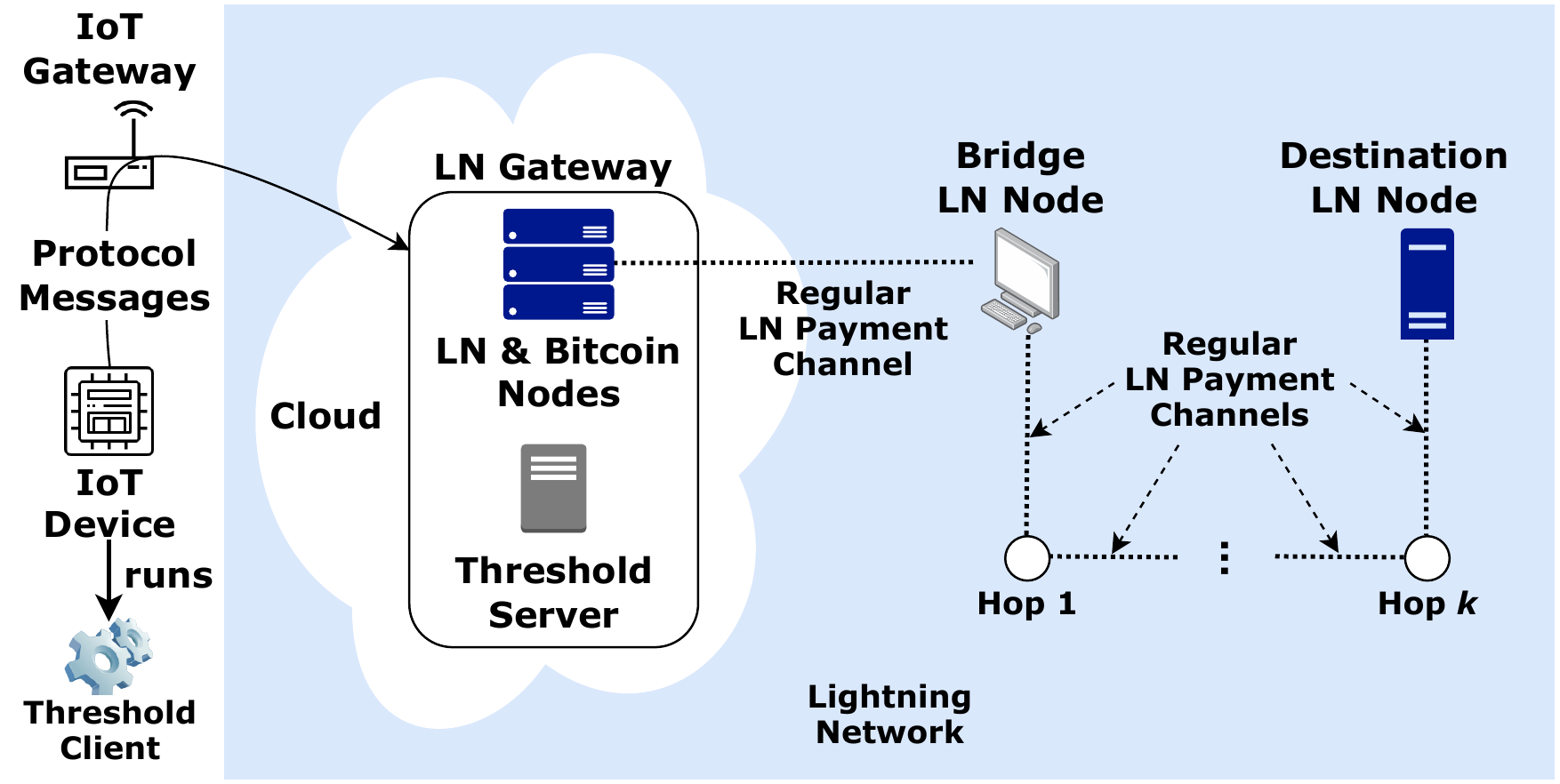}
    \vspace{-7mm}
    \caption{\small Illustration of the system model.}
    \label{fig:system_model}
    \vspace{-2mm}
\end{figure}

There are four main entities in our system which are \textbf{IoT device}, \textbf{LN gateway}, \textbf{Bridge LN node}, and \textbf{Destination LN node} as shown in Fig. \ref{fig:system_model}. We also show other tools and intermediary devices; \textbf{Threshold client}, \textbf{IoT gateway}, \textbf{LN gateway's Bitcoin and LN nodes}, \textbf{Threshold server}. IoT device wants to pay the destination LN node for the goods/services. It is connected to an IoT gateway acting as an access point through a wireless communication standard depending on the IoT application e.g., WiFi, Bluetooth, 5G, etc. The IoT gateway is responsible for connecting the IoT device to the Internet when the IoT device is within its range. Through this internet connection, IoT device can reach the LN gateway, which manages the LN \& Bitcoin nodes and the threshold server. The LN gateway can be hosted anywhere on the cloud and it provides services to the IoT device by running the required full Bitcoin and LN nodes and is \textit{incentivized by the fees IoT device pays in return}. The LN gateway also runs a threshold server that communicates with the threshold client installed at the IoT device when required. This client/server setup enables the 2-party threshold ECDSA operations. Bridge LN node is the node to which the LN gateway opens a channel when requested by the IoT device. This node could be any LN node on the Internet and is determined by the LN gateway. Bridge LN node may charge a routing fee for payments initiated by the LN gateway which creates an incentive for it to accept the channel opening requests coming from the LN gateway. Through the bridge LN node, IoT device's payments are routed to a destination LN node specified by the IoT device. Our proposed protocol requires changes to the LN protocol. However, only the LN gateway has to run the modified LN software and the bridge LN node can use the original LN software as the threshold operations are only between the LN gateway and the IoT device. After they successfully generate the proper ECDSA signatures, it is no different than as if the LN gateway generated this signature by itself using the original LN software. This feature of the proposed system makes the protocol \textit{compatible with the existing LN nodes}.

We assume that IoT device and the LN gateway do not go offline in the middle of a process such as sending a payment. IoT device can be offline for the rest of the time.

\subsection{Threat Model}
\label{sec:threatmodel}

We make the following security related assumptions:

\begin{itemize}[leftmargin=*]

    \item The LN gateway is malicious and it can post old (revoked) channel states to the blockchain. It can gather information about IoT devices and can try to steal/spend funds in the channel as well as deviate from the protocol specifications. 

    \item We assume that any two parties can collude with each other (i.e. the LN gateway and bridge LN node collusion).

    \item We assume that the communication between the IoT device and the LN gateway is encrypted and authenticated (i.e. with AES-256 and HMAC).

\end{itemize}

Handling the revoked state broadcasts is explained in the protocol details rather than the security analysis as we propose modifications to the LN protocol to handle them. We consider the following attacks to our proposed system:

\begin{itemize}[leftmargin=*]
    
    \item \textbf{Threat 1: Collusion Attacks:} The LN gateway and the bridge LN node can collude against the IoT device. Similarly, the IoT device and the LN gateway can collude against the bridge LN node. Lastly, the bridge LN node and IoT device can collude against the LN gateway. The main motivation of all these collusions is to steal money from the third entity. 
     
    \vspace{1mm}
    
    \item \textbf{Threat 2: Stealing IoT Device's Funds:} The LN gateway can steal IoT device's funds that are committed to the channel by 1) sending them to other LN nodes; 2) broadcasting revoked states or; 3) colluding with the bridge LN node.

    \vspace{1mm}

    \item \textbf{Threat 3: Ransom Attacks:} The LN gateway can deviate from the protocol after opening a channel for the IoT device and not execute IoT device's requests (i.e. uncooperative LN gateway). Then, it can ask the IoT device to pay a ransom before executing the payment sending or channel closing requests.
    
\end{itemize}

\vspace{-2mm}
\section{Proposed Protocol Details}
\label{sec:protocol}

This section explains the details of the protocol that includes the channel opening, sending a payment, and channel closing. As mentioned in Section \ref{sec:bolts}, we have modified LN's BOLT \#2 and these modifications are explained throughout the protocol descriptions.

\vspace{-2mm}
\subsection{Channel Opening Process}
\label{sec:channelopening}

The IoT device is not able to open a channel by itself as it does not have access to LN nor Bitcoin network therefore, we enable the IoT device to securely initiate the channel opening process through the LN gateway and jointly generate signatures with it. This means that LN's current channel opening protocol needs to be modified according to our needs. Specifically, we propose to utilize a (2,2)-threshold scheme between the IoT device and the LN gateway for LN operations. Using a threshold scheme in LN operations is the main novelty of our approach. 

All the steps for the channel opening protocol which includes the default LN messages and our additions are depicted in Fig. \ref{fig:openchannel}. We explain the protocol step by step below:

\begin{figure}[h]
    \centering
    \vspace{-1mm}
    \includegraphics[width=\linewidth]{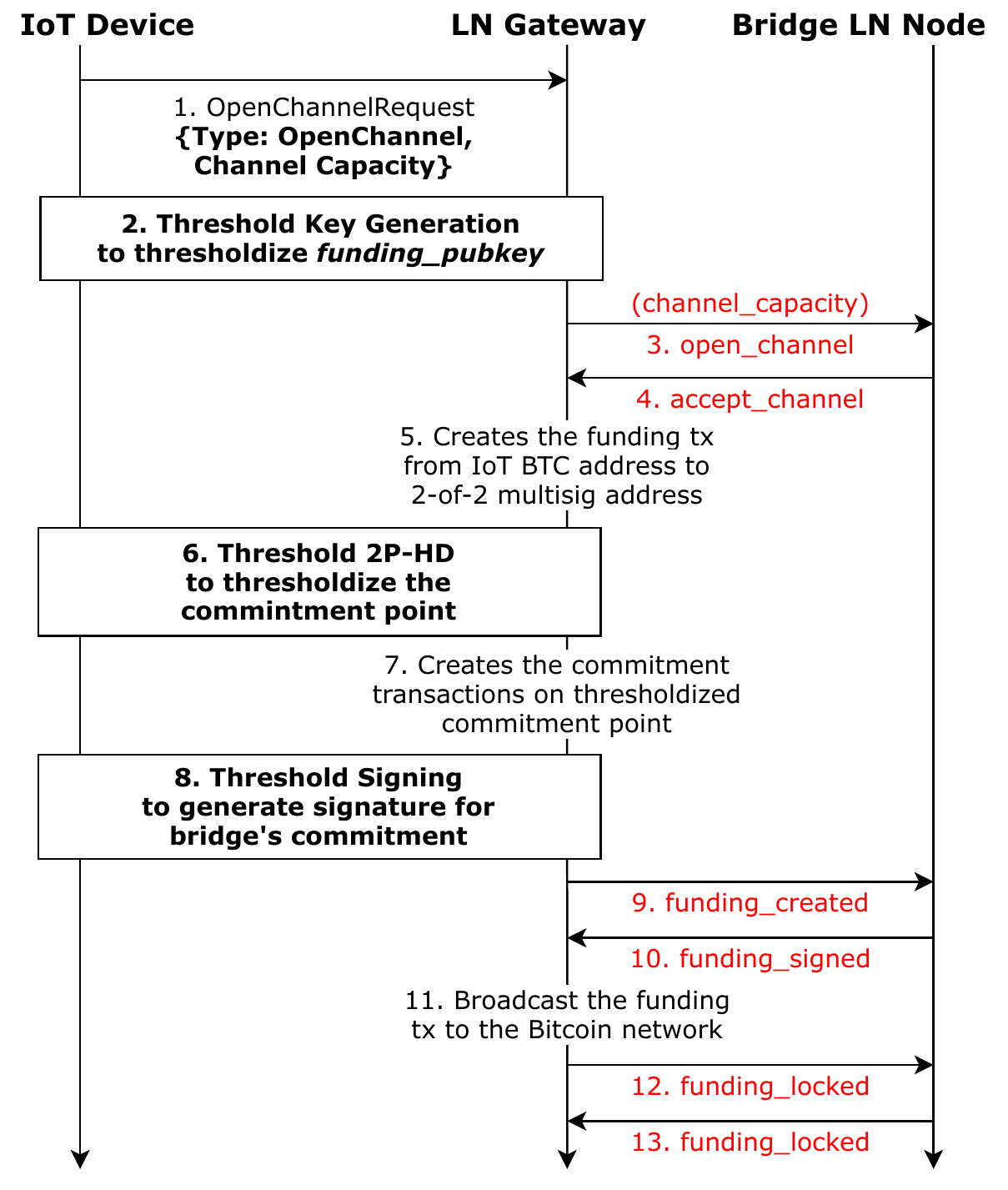}
    \vspace{-7mm}
    \caption{\small Protocol steps for opening a channel. Messages in red show the default messages in BOLT \#2.}
    \label{fig:openchannel}
    \vspace{-4mm}
\end{figure}

\begin{itemize}[leftmargin=*]
\item \textbf{IoT Channel Opening Request}: IoT sends an \textit{OpenChannelRequest} message (Message \#1 in Fig. \ref{fig:openchannel}) to the LN gateway to request a payment channel to be opened. This message has the following fields: \textit{Type: OpenChannelRequest, Channel Capacity}. \textit{Channel Capacity} is specified by the IoT device and this amount of Bitcoin is taken from IoT device's Bitcoin wallet as will be explained in the next steps.

\item \textbf{Channel Opening Initiation}: Upon receiving the request from the IoT device, the LN gateway initiates the channel opening process by connecting to a bridge LN node. The LN gateway can connect to a bridge LN node with the highest number of active LN channels to have a high chance of successfully routing IoT device's payments. This node can be found quickly by querying the network. The LN gateway then sends an \textit{open\_channel} message (\#3 in Fig. \ref{fig:openchannel}) to the bridge LN node which includes its \textit{funding\_pubkey}. \textit{funding\_pubkey} is a Bitcoin public key and both channel parties have their own. The \textit{channel capacity} specified by the IoT device is also sent with this message. Here, we propose to \textbf{thresholdize} LN gateway's \textit{funding\_pubkey} by replacing it with a threshold public key that is jointly computed between the IoT and the LN gateway using (2,2)-threshold key generation (\#2 in Fig. \ref{fig:openchannel}). After this step, bridge LN node responds with an \textit{accept\_channel} message (\#4 in Fig. \ref{fig:openchannel}) to acknowledge the channel opening request of the LN gateway. Note that, \textit{open\_channel} and \textit{accept\_channel} are default BOLT \#2 messages.

\item \textbf{Creating the Transactions}: Now that the channel parameters are agreed on, the LN gateway can create a funding transaction from IoT device's Bitcoin address to the 2-of-2 multisignature address of the channel. Since the input to the funding transaction is from IoT device's BTC address, IoT device can also pay the on-chain fee for this transaction\footnote{Custom funding transactions like this can be constructed by first creating a partially signed Bitcoin transaction (PSBT), then adding the inputs externally and finalizing it in LN. See: \url{https://lightning.readthedocs.io/lightning-openchannel_init.7.html}}. At this step, the LN gateway also creates the commitment transaction for itself and the bridge LN node. Here, we propose to compute the first \textit{commitment point} jointly between the IoT device and the LN gateway. Commitment points are used to derive revocation keys and they are unique for each channel state. Thus, thresholdizing the commitment points prevents the IoT device and the LN gateway from single-handedly revealing the revocation key before the channel state is updated. For this, we propose to use a (2,2)-threshold child key derivation process (\#6 in Fig. \ref{fig:openchannel}) also known as, 2P-HD \cite{gothamcitywhitepaper}. 2P-HD allows the derivation of child keys from the master key that was already generated at Step 2 in Fig. \ref{fig:openchannel}. We propose using 2P-HD over (2,2)-threshold key generation because of its efficiency.

\item \textbf{Exchanging Signatures}: Now, the LN gateway needs to send the signature for bridge LN node's version of the commitment transaction to bridge LN node. For this, the LN gateway and the IoT device jointly generates the signature in a (2,2)-threshold signing (\#8 in Fig. \ref{fig:openchannel}). After signing is done, the LN gateway sends the signature to the bridge LN node along with the outpoint of the funding transaction in a \textit{funding\_created} message (\#9 in Fig. \ref{fig:openchannel}). Learning the funding outpoint, the bridge LN node is now able to generate the signature for the LN gateway's version of the commitment transaction and sends it over to the LN gateway in \textit{funding\_signed} message (\#10 in Fig. \ref{fig:openchannel}). 

\item \textbf{Broadcasting the Transaction to the Bitcoin Network}: After the LN gateway receives the \textit{funding\_signed} message from the bridge LN node, it must broadcast the funding transaction to the Bitcoin network (\#11 in Fig. \ref{fig:openchannel}). Then, the LN gateway and bridge LN node should wait for the funding transaction to reach a specified number of confirmations on the blockchain (generally 3 confirmations). After reaching the specified depth, the LN gateway and the bridge LN node exchange \textit{funding\_locked} messages which finalizes the channel opening (\#12-13 in Fig. \ref{fig:openchannel}).

\end{itemize}

\vspace{-2mm}
\subsection{Sending a Payment}

Similar to the channel opening, we incorporate the IoT device in threshold operations to authorize a payment sending. The same threshold schemes are utilized and LN's current payment sending protocol is modified. The steps of the proposed payment sending protocol is depicted in Fig. \ref{fig:sendpayment} and elaborated below:

\begin{figure}[h]
    \centering
    \includegraphics[width=\linewidth]{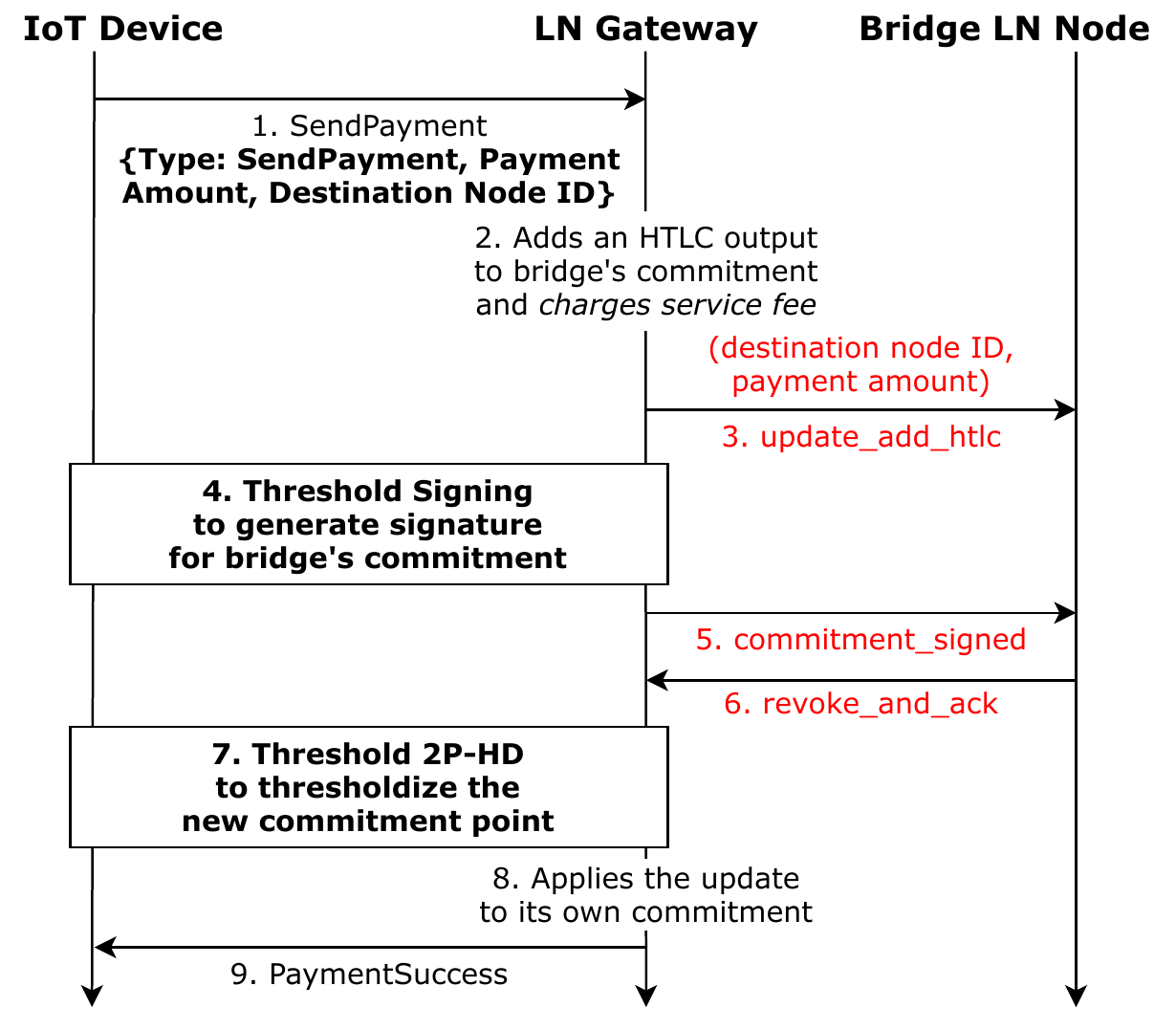}
    \vspace{-8mm}
    \caption{\small Protocol steps for sending a payment. Messages in red show the default messages in BOLT \#2.}
    \label{fig:sendpayment}
    \vspace{-7mm}
\end{figure}

\begin{itemize}[leftmargin=*]

    \item \textbf{Payment Sending Initiation}: To request a payment sending, IoT device sends a \textit{SendPayment} message (\#1 in Fig. \ref{fig:sendpayment}) to the LN gateway. This message has the following fields: \textit{Type: SendPayment, Payment Amount, Destination Node ID}. Here, we assume that \textit{Destination Node ID} is interactively provided to the IoT device in some form (i.e. QR code) by the vendor (i.e. toll gate).

    \item \textbf{Payment Processing at the LN Gateway}: 
    Upon receiving the request, the LN gateway adds an HTLC output to bridge LN node's version of the commitment transaction (\#2 in Fig. \ref{fig:sendpayment}). When preparing the HTLC, the LN gateway \textit{deducts a certain amount of fee} from the real payment amount IoT device wants to send to the destination. Therefore, the remaining Bitcoin is sent with the HTLC. The gateway then sends \textit{update\_add\_htlc} message (\#3 in Fig. \ref{fig:sendpayment}) to the bridge LN node to actually offer the HTLC which can be redeemed in return for a payment preimage. Here, the \textit{destination node ID}, specified by the IoT device, is embedded into the \textit{onion routing packet} which is sent with the \textit{update\_add\_htlc} message.

    \item \textbf{Sending the Signatures}: At this point, the LN gateway can apply the changes to the bridge LN node's commitment transaction. We propose IoT device and the LN gateway to jointly generate a signature for this new commitment transaction in a (2,2)-threshold signing (\#4 in Fig. \ref{fig:sendpayment}). The HTLC signature is also jointly generated at this step. Once the commitment and HTLC signatures are generated, they are sent to the bridge LN node in a \textit{commitment\_signed} message (\#5 in Fig. \ref{fig:sendpayment}). These signatures will enable bridge LN node to spend the new commitment transaction and the HTLC output.

    \item \textbf{Revoking the Old State}: Upon receiving the \textit{commitment\_signed} message, the bridge LN node checks the signatures. Once it verifies that the signatures are valid, it replies with a \textit{revoke\_and\_ack} message (\#6 in Fig. \ref{fig:sendpayment}) which revokes the old state. The LN gateway can use the \textit{commitment secret} included in this message to generate the \textit{revocation private key} which is used in case of revoked state broadcast attempts. Here, we also propose to thresholdize the commitment point for the new channel state between the IoT device and the LN gateway in a 2P-HD (\#7 in Fig. \ref{fig:sendpayment}). Now, it is time for the LN gateway to apply the updates to its own commitment transaction as well (\#8 in Fig. \ref{fig:sendpayment}) which finalizes the payment.

    \item \textbf{Notifying IoT}: Now that the payment sending is finalized, the LN gateway can notify the IoT device of the successful payment by sending a \textit{PaymentSuccess} message (\#9 in Fig. \ref{fig:sendpayment}). If payment sending fails for some reason, the LN gateway can retry in the future without needing to involve the IoT device.
    
\end{itemize}

\vspace{-4mm}
\subsection{Channel Closing Process}

A channel in LN is closed either unilaterally by one of the parties broadcasting its most recent commitment transaction to the blockchain or closed mutually by both parties agreeing on the closing fee. In our case, all 3 parties of the channel namely; the IoT device, the LN gateway, and the bridge LN node can close the channel. We explain all three cases separately below:

\vspace{-1.5mm}
\subsubsection{IoT Device Channel Closure}

When the IoT device would like to close the channel, it follows the proposed protocol below.

1) The IoT device sends a \textit{ChannelClosingRequest} message to the LN gateway to request closing the channel. 

2) The LN gateway has two options to close the channel which are \textit{unilateral} or \textit{mutual} close. In mutual close, the LN gateway and the bridge LN node first exchange \textit{shutdown} messages and then start negotiating on the closing fee. Once they agree on the fee, the closing transaction will be broadcast to the blockchain. We propose this closing transaction to be signed jointly between the IoT device and the LN gateway in a (2,2)-threshold signing before broadcasting. In unilateral close case, the LN gateway just broadcasts its most recent commitment transaction to the Bitcoin network after getting it (2,2)-threshold signed with the IoT device. We propose the on-chain fee for both cases to be paid by the IoT device since the channel closing was requested by the IoT device. The fee is deducted from IoT device's balance in the channel. Once the broadcast transaction is mined, the channel is closed and everyone's funds in the channel settle in their respective Bitcoin addresses.

\vspace{-1.5mm}
\subsubsection{LN Gateway Channel Closure}
\label{sec:gatewaychannelclose}

The LN gateway can also close the channel that was opened due to a request from the IoT device. The steps are very similar to IoT device channel closure case. The proposed protocol steps are as follows:

1) The LN gateway sends a \textit{ChannelClosingRequest} message to the IoT device to show its intention to close the channel. 
    
2) Then, the same steps in 2) in IoT device channel closure are performed. The only difference is that, since now the channel closing is requested by the LN gateway, the on-chain fee is paid by the LN gateway by deducting the fee from its channel balance. Because of this, the LN gateway can only close the channel after collecting enough fees on the channel that can cover the on-chain fee of the transaction. This is the intended behavior anyways because the LN gateway should not close the channel before the IoT device. It should wait for the IoT device to deplete its channel balance first. This way, the LN gateway can maximize the fees it collects on the channel.

\vspace{-2mm}
\subsubsection{Bridge LN Node Channel Closure}

Since bridge LN node runs the original unmodified LN software, its channel closure does not involve any modifications in particular. It can choose to close the channel unilaterally or mutually. We just propose that, as soon as the LN gateway is aware of a channel closure by the bridge LN node, it notifies the IoT device with a \textit{ChannelClosed} message. Upon receiving this message, IoT device will be aware that the channel was closed and it can no longer be used to send payments.

\vspace{-2mm}
\subsection{Handling Revoked State Broadcasts}
\label{sec:handlingrevoked}

In Section \ref{sec:preliminaries}, we explained how Alice and Bob can cheat against each other by broadcasting a \textbf{revoked state} to the blockchain. To address this issue in general, LN uses \textit{timelocks} for the commitment transaction outputs which prevents cheating party to spend the funds immediately. While LN's current scheme secures the nodes against this threat, it needs to be adapted to our case where an IoT device is introduced to the LN channels. Because, in our proposed system, an IoT device cannot monitor the blockchain for broadcast revoked states. It also does not store any previous channel states or their respective revocation private keys. Therefore, essentially, IoT device will not notice when other channel parties try to cheat. On the other hand, the LN gateway and bridge LN node can broadcast revoked states. We examine the two possible cases separately below and propose solutions to handle them to prevent any loss of funds for the IoT device.

\vspace{-1.5mm}
\subsubsection{Revoked State Broadcast by the LN  Gateway}

The LN gateway can broadcast its old (i.e. revoked) commitment transactions to the blockchain in an attempt to cheat against other channel parties. The LN gateway can do that because these revoked commitment transactions are already signed by all channel parties. Thus, when they are broadcast to the Bitcoin network, they will be mined by Bitcoin miners since the transactions are valid. This action has two possible outcomes: 1) Bridge LN node was offline for an extended period of time, therefore did not notice the LN gateway's cheating attempt and loses some or all of its funds depending on the broadcast old channel state. 2) Bridge LN node was online during the LN gateway's cheating attempt, therefore punishes the LN gateway and IoT device by taking all the funds in the channel using the respective revocation private key of the old channel state.

The first scenario does not adversely affect the IoT device. It is bridge LN node's responsibility to stay online to prevent such cheating attempts from the LN gateway. Indeed, all LN nodes are susceptible to this threat when they are offline \cite{offlinelnnodes} thus, this is not a threat specific to our protocol and is beyond the scope of this paper\footnote{Watchtowers \cite{watchtowers} are proposed to remedy this problem. They can monitor an LN node's channels for it and protect them against revoked state broadcast attacks.}.

\begin{figure}[htb]
    \centering
    \vspace{-3mm}
    \includegraphics[width=0.9\linewidth]{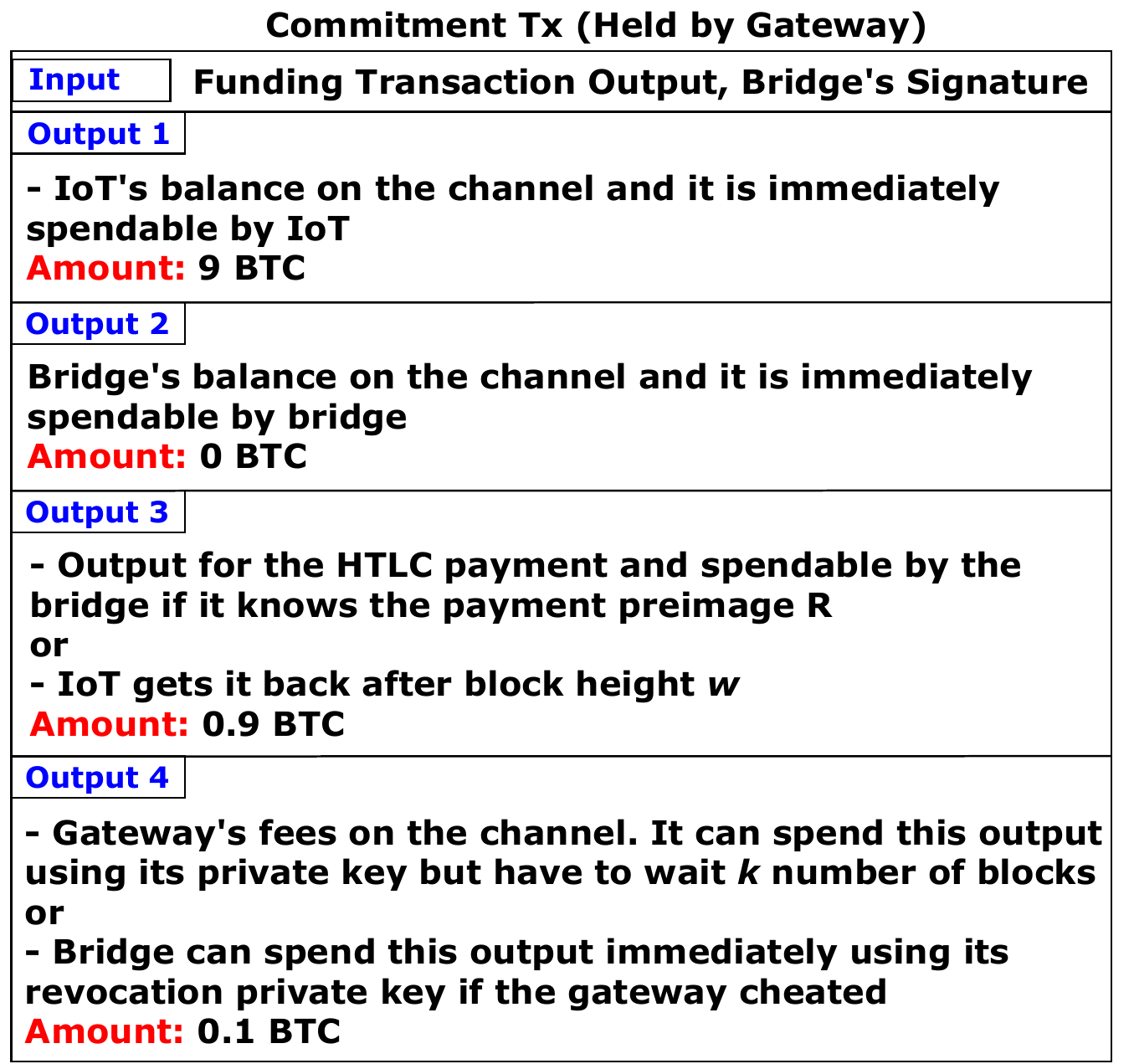}
    \vspace{-3mm}
    \caption{\small Depiction of the proposed commitment transaction for the LN gateway. This commitment transaction is generated after the following operations: 1) A channel with 10 BTC capacity was opened, 2) IoT requested sending 1 BTC to a destination, 3) Gateway charged the IoT a service fee of 0.1 BTC for this payment.}
    \vspace{-4mm}
    \label{fig:gatewaycommitment}
\end{figure}

However, the second scenario will result in loss of funds of the IoT device if not addressed. To circumvent this threat, we propose modifying the LN gateway's commitment transaction outputs such that the LN gateway can get no benefit from cheating. Our proposed commitment transaction for the LN gateway is shown in Fig. \ref{fig:gatewaycommitment}. Basically, the main goal is to prevent IoT device from losing its funds. Therefore, IoT device's balance on the channel shown in Output 1 in Fig. \ref{fig:gatewaycommitment} is protected such that, bridge LN node cannot spend it even when the LN gateway cheats. In other words, this output can only be spent by the IoT device using its Bitcoin private key once the channel is closed. Additionally, we introduce a fourth output to the commitment transaction which is shown as Output 4 in Fig. \ref{fig:gatewaycommitment}. This output is for the LN gateway's fees which it collects from IoT device's payments. We propose to make this output conditional such that, if LN gateway cheats, it risks losing the fees it collected throughout the lifetime of the channel. In other words, the LN gateway's fees will be taken by the bridge LN node if the LN gateway gets caught cheating. With these modifications, the LN gateway is completely disincentivized from cheating and incentivized to follow the protocol specifications.

\vspace{-2mm}
\subsubsection{Revoked State Broadcast by the Bridge LN Node}

Similar to the LN gateway, bridge LN node can also broadcast revoked states to the blockchain. Here, we face the same ``being online" issue for the LN gateway. Basically, if the LN gateway is offline for an extended period of time (i.e. due to a power outage) while bridge LN node is cheating, it will lose some or all the fees it collected on the channel depending on the broadcast old channel state. Therefore, like all existing LN nodes, the LN gateway also has to stay online not to lose any funds. Since this is a general LN problem, we do not propose solutions to fix it, as it is outside the scope of this paper. Additionally, thanks to our proposed Output 1 in Fig. \ref{fig:gatewaycommitment}, this attack does not affect the IoT device's funds in any way.

\vspace{-2mm}
\section{Security Analysis}
\label{sec:securityanalysis}

In this section, we analyze how the presented threats in Section \ref{sec:threatmodel} can be mitigated by our approach. 

\vspace{1mm}
\noindent \textbf{Threat 1: Collusion Attacks}: We explained how our proposed protocol handles the revoked state broadcast attacks. The collusion attacks are similar in the sense that they are initiated by broadcasting a revoked state. We examine all three possible collusion cases below:

\noindent \textit{1) The LN Gateway and Bridge LN Node Collusion}: This can happen in two ways: \newline \indent a) The bridge LN node sends an old state to the blockchain in which it had more funds than it has in the most recent state. The LN gateway does nothing and lets the bridge LN node get away with some extra funds. Later, they split the extra funds between each other. While this is a valid collusion where profit can be made, it is inapplicable to our protocol. Because in our system, payments are always sent from the IoT device to some destination LN nodes which in turn always increases the balance of the bridge LN node on the channel. Thus, bridge LN node has no old states in which it had more funds than it has in the most recent state consequently invalidating the collusion. One-way payments are currently a limitation of the protocol and enabling bidirectional payments for the IoT device was left as a future work. \newline \indent b) The LN gateway broadcasts an old state to the blockchain. As explained in the previous scenario, this is beneficial only if the LN gateway has an old state with more funds. However, this is not the case as the LN gateway charges the IoT device for each new payment which results in its channel balance to always increase. Additionally, the bridge LN node also has less funds in the old states. Since both the LN gateway and the bridge LN node have less funds in old states, this collusion is unprofitable for them.

\noindent \textit{2) IoT Device and the LN Gateway Collusion}: Since IoT device's funds in the channel always decrease with each new payment, it is tempting for it to collude with the LN gateway to broadcast a revoked state to the blockchain together. Specifically, the LN gateway would broadcast the first channel state where the IoT device had all the channel funds and bridge LN node did not have any funds. If bridge LN node cannot act in time to stop this cheating attempt, it would lose all its funds in the channel. However, this takes us back to the ``being online" problem again that we mentioned in Section \ref{sec:handlingrevoked}. If the bridge LN node does not want to lose money, it must not go offline for extended periods of time. Therefore, again, this collusion is not specific to our protocol, but rather a general LN issue.

\noindent \textit{3) IoT Device and the Bridge LN Node Collusion}: This collusion case is unfeasible as the bridge LN node is chosen by the LN gateway at the channel opening stage. Thus, IoT device cannot possibly collude with this entity which it does not know.

\vspace{1mm}
\noindent \textbf{Threat 2: Stealing IoT Device's Funds:} Using (2,2)-threshold signatures for the LN operations secure the IoT device's funds in the channel since the LN gateway: 1) cannot send IoT device's funds in the channel to other LN nodes without generating proper HTLC and commitment signatures with the IoT device in a (2,2)-threshold signing; 2) cannot cause loss of IoT device's funds by broadcasting revoked states as shown in Section \ref{sec:handlingrevoked} and; 3) cannot cause loss of IoT device's funds by colluding with the bridge LN node as shown in Threat 1 above. If we used LN's original signing mechanism, the LN gateway could move IoT device's funds in the channel without needing a signature from the IoT device. Consequently, usage of (2,2)-threshold along with the proposed modifications to the LN gateway's commitment transaction prevent IoT device from losing any funds.

\vspace{1mm}
\noindent \textbf{Threat 3: Ransom Attacks:} 
This is an attack where the LN gateway is deviating from the protocol description. To put it with some examples, the LN gateway can say IoT device that, ``I will close this channel only if you pay me X amount of Bitcoins" or, ``from now on, I will execute your payment sending requests only if you accept to pay an \%10 increased service fee". This essentially turns into a game where IoT device's best move is to reject the ransom attempt and just wait. Then the LN gateway holds IoT device's funds hostage for as long as it can in an attempt to deter the IoT device. This is a \textit{deadlock} case where both parties just wait. It is clear that the LN gateway has no incentive to perform ransom attacks as it does benefit from it assuming that the IoT device acts rationale. The best course of action for the LN gateway is to continue serving IoT device and keep collecting the service fees. Thus, our proposed protocol protects the IoT device against ransom attacks.

\vspace{-3mm}
\section{Evaluation}
\label{sec:evaluation}
This section describes the experiment setup to implement the proposed approach and presents the performance results. 

\vspace{-3mm}
\subsection{Experiment Setup and Metrics}

To evaluate the proposed protocol, we implemented it by modifying the source code of c-lightning \cite{clightning} which is one of the implementations of the LN protocol written in C. The \textit{hsmd} module of c-lightning was modified. Hsmd module manages the cryptographic operations and controls the funds in the channel. Specifically, we thresholdized the \textit{funding\_pubkey} and the associated Bitcoin private key as explained in Section \ref{sec:channelopening}. The LN gateway ran this modified version of c-lightning. To the best of our knowledge, this is the first-ever work that implemented threshold cryptography for LN. Thresholdizing the commitment points is a work-in-progress therefore its overhead is not included in the experiments. Our implementation is publicly available in our GitHub repository at \url{https://github.com/ZenGo-X/lightning}.

Our test setup is as follows: For the IoT device, we used a Raspberry Pi 3 Model B v1.2 and it was located at Florida with a public IP. For the LN gateway, we used an Intel Premium droplet from Digitalocean \cite{digitalocean} with 4 cores, second generation Intel Xeon Scalable processor and 8 GB RAM. The droplet was located at New York with the intention to make it a remote machine for the IoT device. For the threshold client/server application, we used our modified version of gotham-city \cite{gothamcity} which is a minimalist decentralized Bitcoin wallet for 2-party threshold ECDSA written in Rust programming language. For the full Bitcoin node, we used \textit{bitcoind} \cite{bitcoind} which is one of the most widely used implementations of the Bitcoin protocol.

We used IEEE 802.11n (WiFi) to exchange protocol messages between Raspberry Pi and the IoT gateway. An AT\&T BGW320-500 Fiber Internet modem was used for the IoT gateway. We created a server \& client socket application in Python and connected Raspberry Pi to the IoT gateway through the created TCP socket in the local WiFi network. The IoT gateway then forwarded the protocol messages to the LN gateway running on the cloud. 

To assess the performance of our protocol, we used the following metrics: 1) \textit{Time} which refers to the total communication and computational delays of the proposed protocol; 2) \textit{Cost} is the total monetary cost associated with opening channels, sending payments, and closing channels using the LN gateway; 3) \textit{Network} which refers to the network usage of the IoT device and the required bandwidth (data rate) for the IoT device for timely LN operations.

To compare our approach to a baseline, we considered the case where the LN gateway performs the LN operations by itself such as sending a payment. In other words, no IoT device is present, and all LN tasks are solely performed by the LN gateway. We will refer to it as \textit{no IoT case} in next sections.

\vspace{-4mm}
\subsection{Communication and Computational Delays}
\label{sec:overheadresults}

We first assessed the computational and communication delays of our proposed protocol. Computational overhead of running the protocol on the Raspberry Pi involves 3 different computations. These are the AES encryption of the protocol messages, HMAC calculations and threshold computations. We used Python's \textit{pycrypto} library to encrypt the protocol messages with AES-256 encryption. The encrypted data size for the messages was 24 bytes. For the authentication of the messages, we used HMAC. To calculate the HMACs, \textit{hmac module} in Python was used. As mentioned before, for the threshold operations, we used our modified version of gotham-city \cite{gothamcity}. Gotham-city uses a REST API to relay the messages between the threshold parties (i.e. the client and the server). All values are mean for 10 runs of the respective operation. The presented threshold key generation and signing times were measured using localhost for the server. In this way, there were no communication delays involved and pure computation times were measured. We present these computation times at Table \ref{tab:computational}.

\begin{table}[h]
  \begin{center}
  \vspace{-3mm}
    \caption{Computation Times on the IoT Device}
     \vspace{-3mm}
    \label{tab:computational}
    \resizebox{0.9\linewidth}{!}{
    \begin{tabular}{|c|c|c|c|}
    \hline
       \textbf{AES}        &  \textbf{HMAC}  & \textbf{Threshold Key} &  \textbf{Threshold}   \\
       \textbf{Encryption} & \textbf{Calculation} & \textbf{Generation} &  \textbf{Signing}  \\ 
       \hline
               15 ms        &       $<$ 1 ms         & 1.05 s  &  153 ms  \\
     \hline
    \end{tabular}
    }
   \vspace{-2mm}
  \end{center}
\end{table}

As can be seen from the results, the overhead of the AES encryption and HMAC calculation is negligible. Threshold key generation takes about 1 second but it is not critical for LN operations as it is done 1-time at channel opening as explained in Section \ref{sec:channelopening}. The threshold signing time is only 153 ms which is again almost negligible.

We then measured the communication delays which are used in other experiments to evaluate the timeliness of the protocol. Communication delay has the following components: 1) The delay of sending a request from the IoT device to the LN gateway; 2) Execution time of the respective LN operation at the LN gateway including the execution time of the threshold operation between the IoT device and the LN gateway; and 3) The delay of the acknowledgment message sent from the LN gateway to the IoT device. For these delays to be comparable, we also measured the \textit{no IoT case} delays where the LN gateway performs the LN operations by itself. In this case, there are no threshold operations and no communication with the IoT device. We present the delays for these two cases in Table \ref{tab:communication}.

\begin{table}[h]
  \begin{center}
  \vspace{-2mm}
    \caption{Communication Delays of Various Operations for ``No IoT Case" and ``IoT Case"}
     \vspace{-3mm}
    \label{tab:communication}
    \resizebox{\linewidth}{!}{
    \begin{tabular}{|c|c|c|c|c|c|c|}
    \hline
                & \textbf{IoT}       & \textbf{Connecting} & \textbf{Channel} &  \textbf{Payment}  &  \textbf{Channel} & \textbf{Ack}  \\
                & \textbf{Request}   & \textbf{to Bridge}  & \textbf{Opening} & \textbf{Sending}   &  \textbf{Closing} & \textbf{Sending} \\ 
                & \textbf{Time}      & \textbf{LN Node}    & \textbf{Time}    & \textbf{Time}      &  \textbf{Time}    & \textbf{Time} \\ 
       \hline
          \textbf{IoT}  &  55 ms  & 6.22 s  & 3.86 s     &  4.120 s       &  4.005 s         & 46 ms \\
          \textbf{Case}  &  &   &      &      &          &  \\ \hline
          \textbf{No IoT}  &  -  & 0.403 s  & 0.267 s     &  1.462 s       &  0.737 s         & - \\
          \textbf{Case}  &    &   &    &      &     &  \\  \hline
    \end{tabular}
    }
   \vspace{-3mm}
  \end{center}
\end{table}

As can be seen from the Table \ref{tab:communication}, no IoT case is quicker compared to the IoT case, as expected. The difference in delays is around 3 seconds for most LN operations. This difference in communication delays between the two cases is introduced by the threshold operations performed between the IoT device and the LN gateway. It was already shown in Table \ref{tab:computational} that, the computational times for threshold operations were not significant. This tells us that, most of the time for threshold operations are just communication delays of the exchanged messages between the IoT device and the LN gateway. Note that, the channel opening and channel closing times do not include the confirmation times on the blockchain.

\vspace{-2mm}
\subsection{Cost Analysis}
We consider and analyze the following costs associated with our proposed protocol: 1) The on-chain fees for channel opening and closing; 2) The fees that are charged by the LN gateway for IoT device's payments; 3) Forwarding fees that are charged by the nodes on a payment route (i.e. bridge LN node).

According to our experiment results; a channel opening transaction signed with (2,2)-threshold ECDSA, incurs a fee of 222 satoshi when it is desired for the transaction to be included in the next block\footnote{See this channel opening transaction which is signed using (2,2)-threshold ECDSA: \href{https://blockstream.info/testnet/tx/20739b2f5ce19e806a8b1fbe151924c643c8abb28cd635e372d0d7c253a09a57}{20739b2f5ce19e806a8b1fbe151924c643c8abb28cd635e372d0d7c253a09a57}}. Similarly, a unilateral channel close signed with (2,2)-threshold ECDSA, costs 183 satoshi\footnote{See this unilateral channel closing transaction signed with (2,2)-threshold ECDSA: \href{https://blockstream.info/testnet/tx/f23bb92a6decb6bd05eff5907af30294bf0025ef776d53dad906b8d7950e9395}{f23bb92a6decb6bd05eff5907af30294bf0025ef776d53dad906b8d7950e9395}}. At current Bitcoin price of \$54,500, these fees correspond to \textit{12 cents} and \textit{10 cents} respectively which is basically negligible considering these operations are 1-time.

LN gateway's service fee for payments entirely depends on the LN gateway's choice. In Section \ref{sec:tollusecase}, we simulate this cost for a toll payment scenario. Forwarding fees again entirely depend on nodes' choices which are on a payment path. In LN, nodes charge a fixed fee each time they route a payment which is called the \textit{base fee} \cite{lnroutingfees}. There is also \textit{fee per satoshi} that nodes charge proportional to the satoshi value of the payment they route \cite{lnroutingfees}. To measure an approximate value for the forwarding fee, we sent 10 LN payments using different routes for each. The mean of the forwarding fees was \textit{2 satoshi} which is again negligible.

Most importantly; the channel opening, channel closing, and forwarding fees do not change with the introduction of (2,2)-threshold ECDSA. These values we measured are exactly the same for a regular LN node running the original LN software. Therefore, our protocol does not bring extra cost overhead for LN operations.

\vspace{-2mm}
\subsection{Network Usage and Bandwidth Analysis}
In this experiment, we investigate: 1) the \textit{network usage} between the IoT device and the LN gateway; 2) the \textit{required bandwidth} for the IoT device to timely complete the LN operations with the LN gateway.
To measure the network usage, we used the command-line network monitoring tool \textit{IPTraf-ng} \cite{IPTrafng}. This tool enables inspecting the IP traffic of a device. Since the IP address of the LN gateway is known, we can easily learn the number of bytes exchanged with that IP address on the IoT device. We considered all main LN operations which are: 1) Connecting to a bridge LN node, 2) Opening a channel, 3) Sending a payment, 4) Closing a channel. All LN operations were performed 10 times and the mean values were calculated. Our measurement results indicate that the network usage on the IoT device for these LN operations ranges between (3988, 4032) bytes where 30 network packets are exchanged for each. 

Now, the required bandwidth for the IoT device can be calculated.
Considering a payment sending; where the IoT device exchange around 4000 bytes in 4.12 seconds (Table \ref{tab:communication}, Column 4), then the required bandwidth is 7.8 kbps. This is a very reasonable bandwidth requirement considering that even the very low bandwidth LoRa supports up to 27 kbps \cite{adelantado2017understanding}.

\vspace{-2mm}
\subsection{Toll Payment Use-Case Evaluation}
\label{sec:tollusecase}

Real-time response is critical in a toll application where cars pass through a toll gate and pay the toll without stopping by utilizing wireless technologies. When a car enters the communication range of the toll's wireless system, it sends a request to the LN gateway of the toll gate through the IoT gateway to initiate the payment. The LN gateway immediately sends the requested amount of payment to the toll company's LN node. After payment is successfully sent, a \textit{PaymentSuccess} message is sent back to the car through the IoT gateway. For this to work, whole payment sending process must be completed before the car gets out of the communication range of the toll gate's wireless system.

We know from Table \ref{tab:communication} that, payment sending with our proposed protocol takes 4.12 seconds. Assuming that the toll gate uses IEEE 802.11n for the wireless communication, the communication range then will be 250 meters which is the advertised range of 802.11n \cite{802.11nrange}. If cars pass through the toll gate with a speed of 80 miles per hour, there is around 7 seconds available for them to complete the protocol message exchanges with the LN gateway for a successful toll payment. Since the payment sending only takes 4.12 seconds, cars will be able to pay the toll on time.

We now investigate the cost of using this service for a car for 1 month where it makes 2 toll payment in a day for the sake of a complete analysis of the use-case scenario. Assuming that a channel was already opened for the car, the only cost of using the service will be the LN gateway's service fees for each toll payment. While this service fee is decided by the LN gateway, let us assume that it is \%5 on top of the actual toll amount. If toll amount is \$0.75 per pass, then the service fee is \$0.0375 per payment. In 1 month, the car will make 60 toll payments, so the total amount of service fees the LN gateway will be charging is only \$2.25. Considering the convenience of making near-instant toll payments directly from your car, we believe paying an extra \$2.25 in a month would be insignificant to the customers.

\vspace{-4mm}
\section{Related Work}
\label{sec:RelatedWork}

As Bitcoin penetrated into our lives, some recent studies started looking at the integration of it with IoT applications. While some of these studies focus on Bitcoin, others try to integrate LN with IoT. Hannon et al. in \cite{hannon2019bitcoin} propose a protocol based on LN to give the IoT devices the ability to transact on LN. They propose using two untrusted third parties which are called \textit{IoT payment gateway} and \textit{watchdog} respectively. IoT payment gateway posts the transactions to the blockchain and watchdog monitors the gateway to detect cheating and inform IoT. However, their approach has a fundamental flaw. They assumed that the IoT device can open payment channels to the gateway and it is left unclear how IoT can open such channels. This is indeed the exact problem we are trying to solve. IoT devices do not have the computational resources to open and maintain LN payment channels. Therefore, their assumption is not feasible and our work is critical in this sense to fill this gap.

Another work, \cite{pouraghily2019lightweight}, proposes a ticket-based verification protocol with the aim of enabling IoT devices to perform economic transactions. Specifically, they propose to use \textit{contract manager} and \textit{transaction verifier} to cut the processing burden on the IoT devices. However, this approach has major problems: The authors mention a joint account which is opened with a partner device that holds IoT device's funds. This raises security concerns as the details of the joint account is not provided. Additionally, the approach was compared with $\mu$Raiden \cite{raiden} which is an Ethereum based payment channel network. However, the reliability of this comparison is questionable as $\mu$Raiden development was stopped 2 years ago. In contrast to this work, we targeted Bitcoin's LN as it is currently the most actively developed PCN.

Finally, the authors of \cite{robert2020enhanced} propose a module to integrate LN into an existing IoT ecosystem. In their approach, an \textit{LN module} is introduced which has the full Bitcoin and LN nodes to allow data-consumers to access selected items through payments. In contrast to our work, this approach is focusing on integrating LN payments into an existing IoT marketplace. Thus, the individual devices that are not a part of such marketplaces are not considered. Additionally, the authors' LN framework relies on Bitcoin wallets that are held by third parties in the IoT marketplace which raises security and privacy concerns. In our approach, IoT devices do not share the ownership of their Bitcoins with a third party.

In our previous study, we proposed a protocol to enable IoT micro-payments using 3-of-3 multisignature LN channels \cite{kurt2020enabling}. In this approach, LN's original 2-of-2 multisignature LN channels were proposed to be changed as 3-of-3 multisignature. In this way; the IoT device, the LN gateway and the bridge LN node can cooperatively perform LN operations by incorporating their signatures. Our proposed threshold approach in this work does not require modifications on the channel structure which makes it compatible with the rest of the network.

In addition to these research efforts, there have been implementation efforts to create light versions of LN for devices with constrained resources. For instance, Neutrino \cite{lightclient} is one of the Bitcoin \textit{light client} implementations specifically designed for LN. It works based on synchronizing only the block headers and filters instead of the whole blockchain. While a portion of the IoT devices have enough resources to run such light clients, they still have to download the block headers and filters to synchronize which is not realistic in many real-life applications. Therefore, we opt for a solution that will exclude such options and appeal to a wide range of IoT devices and applications.

\section{Conclusion}
\label{sec:conclusion}
In this paper, we proposed a secure and efficient protocol for enabling IoT devices to use Bitcoin's LN for sending payments. By introducing (2,2)-threshold scheme to LN and modifying LN's existing peer protocol for channel management, a third peer (i.e. IoT device) was added to the LN channels. The purpose was to enable resource-constrained IoT devices that normally cannot interact with LN to interact with it and perform micro-payment transactions with other users. IoT device's interactions with LN are achieved through a gateway node that has access to LN and thus can provide LN services to it in return for a fee. In order to prevent possible threats that might arise from cheating or colluding, gateway's Bitcoin scripts were modified. Our evaluation results showed that the proposed protocol enables timely payments for IoT devices with negligible delay/cost overheads.

\begin{acks}
We thank Christian Decker \href{https://twitter.com/Snyke}{(@Snyke)} and Lisa Neigut \href{https://twitter.com/niftynei}{(@niftynei)} for their inputs during the implementation efforts. This work was partially funded by the Cisco Foundation.
\end{acks}

\bibliographystyle{ACM-Reference-Format}
\bibliography{references}

\appendix

\section{BOLT \#2: Peer Protocol for Channel Management}
\label{sec:appendixbolt2}

\noindent \textit{Channel Establishment}: Channel establishment is the first phase of LN's BOLT \#2 and is used to open channels to other LN nodes. There are several message exchanges between the funding node and the responding node (fundee). These messages are: \textit{open\_channel}, \textit{accept\_channel, funding\_created, funding\_signed, funding\_locked}. When these messages are properly exchanged and both nodes follow the protocol steps correctly, channel is opened.

\vspace{1mm}

\noindent \textit{Normal Operation}: Once the channel is opened after the channel establishment phase, nodes can use the channel to send and receive payments via HTLCs. Several messages are exchanged when nodes want to send a payment to the others. These messages are \textit{update\_add\_htlc, update\_fulfill\_htlc, update\_fail\_htlc, update\_fail\_malformed\_htlc, commitment\_signed, revoke\_and\_ack}. Either party of the channel can send \textit{update\_add\_htlc} to offer an HTLC to the other, which can be redeemed with the payment preimage.

\vspace{1mm}

\noindent \textit{Channel Close}: Both parties can close the channel whenever they want. There are two types of channel closing methods in LN, namely \textit{mutual close} and \textit{unilateral close}. In the mutual close case, nodes try to agree on a channel closing fee. Once they agree on the fee and close the channel, they can access their funds immediately. In the unilateral close, however, one of the nodes broadcasts its most recent commitment transaction to the blockchain and closes the channel. The unilateral close is not the preferred way in LN to close a channel because the node might broadcast a revoked commitment transaction which can result in loss of funds of the other node if it is not online. A channel can be closed once all the pending HTLCs in the channel are cleared. Channel closing phase has also some message exchanges and these messages are \textit{shutdown} and \textit{closing\_signed}.

The full details of the protocol can be found at \cite{bolt2}.

\section{(2,2)-Threshold Key Generation}
\label{appendix:keygeneration}

Let $\mathbb G$ be an elliptic curve of prime order $q$ and a base point $G$. We assume that the protocol is run between a server $P_1$ and a client $P_2$.

\begin{enumerate}[leftmargin=*]
    \item Initiation: Choosing a random $x_1$, the server computes $Q_1 = x_1 \cdot G$ and choosing a random $x_2$, the client computes $Q_2 = x_2 \cdot G$.
    
    \item A variation of the ECDH key exchange from \cite{lindell2017fast} is run by the client and the server to generate the joint public key $Q = x_1x_2G$.
    
    \item A Paillier key-pair is generated and $c_{key} = Enc_{pk}(x_1)$ is computed by the server where $pk$ is the Paillier public key. Then, the server sends $c_{key}$ and $pk$ to the client.
    
    \item The client receives a non-interactive zero-knowledge proof whose Paillier key is well-formed from the server. Proof in \cite{cryptoeprint:2018:057} as well as the parameters in \cite{lindell2018fast} were utilized.
    
    \item The client asks the server to prove that the encrypted value in $c_{key}$ is the discrete log of $Q_1$ using a 2-round zero-knowledge protocol that is given at Section 6 of Lindell's work \cite{lindell2017fast}.
    
    \item The client asks the server to prove $x_1 \in	\mathbb Z_q$ where $q$ is the order of the elliptic curve using a 2-round zero-knowledge protocol that is given at Appendix A of Lindell's paper \cite{lindell2017fast}.
\end{enumerate}

\section{(2,2)-Threshold Signing}
\label{appendix:signing}

A two-party ECDSA signing algorithm that will produce the same signature $(r,s)$ as the signature scheme given in Section \ref{sec:ECDSA} is presented below. It was adopted from \cite{gothamcitywhitepaper}.

\begin{enumerate}[leftmargin=*]
    \item Step 1 of the key generation protocol is repeated for ephemeral key-pairs by the server and the client. This outputs $k_1, R_1$ for the server and $k_2, R_2$ for the client.
    
    \item Server and the client can extract the $x$-coordinate $r=r_x$ from the same $R$ they generate by $R=k_1 \cdot R_2$ and $R=k_2 \cdot R_1$ respectively.
    
    \item $c_1 = Enc_{pk}(k_2^{-1} \cdot H(m) + \rho q)$ and $c_2 = {c_{key}}^{x_2 \cdot r \cdot k_2^{-1}}$ where $\rho$ is some random number, are computed by the client. Then, $c_3 = c_1 \oplus c_2$ is computed and sent to the server by the client where $\oplus$ is the additive homomorphic operation of Paillier cryptosystem. 
    
    \item $c_3$ is decrypted by the server to get $s'$ which is used to compute $s = s' \cdot k_1^{-1}$. Now, server outputs $(r,s)$ as the ECDSA signature if $(r,s)$ validates as a signature on $H(m)$.
\end{enumerate}

\end{document}